\documentclass[%
reprint,
 amsmath,amssymb,
 aps,pra
]{revtex4-1}

\usepackage{graphicx}% Include figure files
\usepackage{dcolumn}% Align table columns on decimal point
\usepackage{bm}% bold math
\usepackage[colorlinks=true, allcolors=blue]{hyperref}
\begin{document}

\preprint{APS/123-QED}

\title{Long-range interaction of hydrogen atoms at finite temperatures}% Force line breaks with \\
\author{T. Zalialiutdinov$^{1,2}$}
\email[E-mail:]{t.zalialiutdinov@spbu.ru}
\author{D. Solovyev$^{1,2}$}
\affiliation{ 
$^1$Department of Physics, St. Petersburg State University, Petrodvorets, Oulianovskaya 1, 198504, St. Petersburg, Russia}
\affiliation{ $^2$Petersburg Nuclear Physics Institute named by B.P. Konstantinov of National Research Centre 'Kurchatov Institut', St. Petersburg, Gatchina, 188300, Russia
}

\date{\today}% It is always \today, today,

\begin{abstract}
In this study, we reexamine the long-range interaction between two atoms placed in an equilibrium thermal radiation environment. Employing the formalism of quantum electrodynamics at finite temperatures, we derive an expression for the thermal correction to the interaction potential and explore various asymptotic behaviors. The numerical calculations of temperature-dependent dispersion coefficients for both the ground and highly excited states of the hydrogen atom are performed. We proceed from the first principles of the theory to derive the dipole-dipole interaction at finite temperature. The analysis presented in this work reveals that the expressions established earlier in the context of phenomenological extrapolation from zero- to finite-temperature scenarios exhibit disparate asymptotic behavior and lead to overestimated results to those of the rigorous quantum electrodynamics approach.
%Our analysis reveals that the expressions previously established in the literature, which were derived within the context of a phenomenological extrapolation from zero-temperature results for dipole-dipole interactions to finite-temperature scenarios, exhibit disparate asymptotic behavior and lead to overestimated outcomes in comparison with the rigorous quantum electrodynamical approach.
\end{abstract}

\maketitle

\section{Introduction}

The long-range forces between two stationary atoms or molecules characterized by polarizabilities were initially explored in pioneering studies by Casimir and Polder \cite{CasimirPolder:1948}. In 1956, Lifshitz was arguably among the first to contemplate induced forces between two dipoles at non-zero temperatures~\cite{Lifshitz}. Since then, sporadic yet enduring interest has been exhibited in dipole-dipole interactions at finite temperatures~\cite{Chiu:1985, Gorza2006, Fujii:2022}. Over the past decades various approaches to the theoretical description of this phenomenon have been considered \cite{Dzyaloshinskii1961, Berman:2014, Safari:2015}. Furthermore, recent experimental investigations of interactions among atoms at long distances exposed to heated environment have sparked additional interest in this problem~\cite{Cornell:2007, PasseratdeSilans2014}.

Typically, the transition to finite temperatures is accomplished by a phenomenological generalization of the well-established result at $T=0$ through the introduction of a relevant induced term into the expression for the interaction potential. In the scenario where two interacting atoms are placed within an equilibrium thermal radiation field described by the Planck distribution, this generalization results in the replacement of the term characterizing the vacuum zero-point expectation energy, represented by the term $1/2$, onto $1/2 + n_{\beta}(\omega)$. Here $n_{\beta}(\omega)$ is the Bose-Einstein frequency distribution defined as $n_{\beta}(\omega) = (\exp(\omega/(k_B T))-1)^{-1}$, where $k_B$ represents the Boltzmann constant and $T$ is the temperature in Kelvin. This approach has been employed, notably, in the studies~\cite{Ninham:1998, Goedecke:1999, Passante:2007}.

%However, a comprehensive derivation of the interaction potential between two atoms within the formalism of quantum electrodynamics (QED) at finite temperatures has hitherto been limited to the study presented in~\cite{Chiu:1985}, which primarily offered parametric estimates devoid of specific calculations for particular systems.
However, a comprehensive derivation of the interaction potential between two atoms within the formalism of quantum electrodynamics at finite temperatures (TQED) has hitherto been limited to the study presented in~\cite{Chiu:1985}. This work primarily offers parametric estimates devoid of specific calculations for particular systems. It is pertinent to highlight that the computations conducted within the mentioned investigation do not readily enable a direct comparison between the TQED approach and the phenomenological extension applied to thermal scenarios as witnessed in works~\cite{Ninham:1998, Goedecke:1999, Passante:2007}.

In the present paper, within the framework of TQED utilizing the real-time formalism, we reexamine the derivation of the interaction potential between two atoms at long distances and compare it with findings from prior research. Specifically, we perform a numerical computation of thermal corrections to the dispersion coefficients for two interacting hydrogen atoms in excited states. Calculations are carried out for various asymptotics of interatomic distances and temperature regimes.

The paper is organized as follows. In section \ref{section1} we derive the long-range interaction potential between two atoms within the $S$-matrix formalism and discuss its generalization to the finite temperature case. %Modifying the expression for case when both atoms are of the same type but in different states is also discussed there.
A modification of the resulting expression for two identical atoms, but in different states, is also discussed there. In sections \ref{section2} and \ref{section3} we consider the short and long-range limit of the obtained potential, respectively. Details of numerical calculations with analysis of results are presented in section \ref{final_results}. All additional algebraic calculations are located in the appendixes \ref{appendix:A} and \ref{appendix:B}. The relativistic units (r.u.) are used throughout the paper $\hbar=c=m=1$ ($\hbar$ is the Planck constant, $m$ is the electron mass and $c$ is the speed of light), in which the fine structure constant can be expressed in terms of the electron charge as $\alpha = e^2$. The Boltzmann constant in these units is $k_{B}=m\alpha^2k_{B}^{\mathrm{a.u.}}$, where $k_{B}^{\mathrm{a.u.}} = 3.16681\times 10^{-6}$ is given in atomic units.

\section{Long-range interaction between two atoms: $S$-matrix approach}
\label{section1}

\begin{figure}
    \centering
    \includegraphics[scale=0.6]{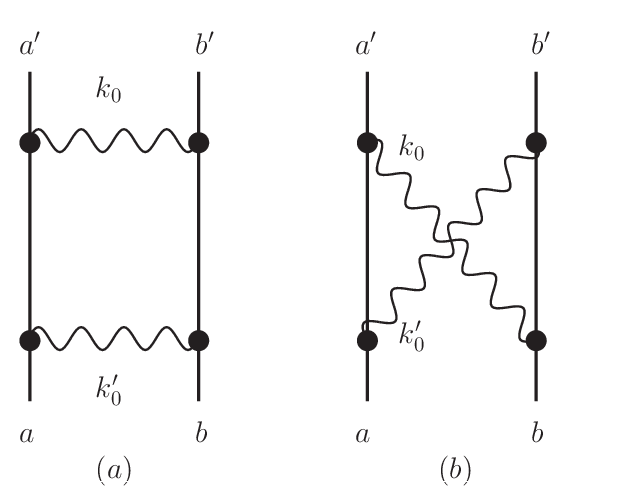}
    \caption{Ladder (a) and crossed-ladder (b) Feynman diagrams describing the long-range interaction of two atoms. Each solid line denotes the particular atom while wavy lines correspond to the exchange of photon. The states of atoms are denoted by $a$, $a'$, $b$, $b'$ for atoms $A$ and $B$, respectively, the photon frequencies are denoted by $k_0$, $k'_0$.}
    \label{fig:1}
\end{figure}

Within the framework of Quantum Electrodynamics (QED) and perturbation theory, the interaction between two atoms, designated as $A$ and $B$, is described by the fourth-order $S$-matrix. The complete set of Feynman diagrams is shown in Fig.~\ref{fig:1}.

We restrict ourselves to the description of the interaction between two one-electron atoms as the main application of the approach developed below. A generalization to many-electron atomic systems can be made in the framework of the general theory, see, e.g., \cite{lindgren}. 
Then, for the ladder (L) diagram, Fig.~\ref{fig:1} (a), the corresponding $S$-matrix element can be written as follows:
\begin{eqnarray}
    \label{s1}
    S_{AB}^{(4), \,L}=(-ie)^4\int dx_1 dx_2 dx_3 dx_4\overline{\psi}_{a'}^{A}(x_1)\overline{\psi}_{b'}^{B}(x_3)
    \\\nonumber
    \times
   \gamma^{\mu_1}_{A} D_{\mu_1\mu_3}(x_1,x_3)\gamma^{\mu_3}_{B}S_{A}(x_1,x_2)\gamma^{\mu_2}_{A}
    \\\nonumber
    \times
    D_{\mu_2\mu_4}(x_2,x_4)
    \gamma^{\mu_4}_{B}
    S_{B}(x_3,x_4)\psi_{a}^{A}(x_2)\psi_{b}^{B}(x_4)
    ,
\end{eqnarray}
where $\psi_{a}^{A}(x)=e^{-i\varepsilon^{A}_{a}t}\psi(\bm{x})$ is the solution of Dirac equation for bound electron in the state $a$ of the atom $A$, $\overline{\psi}_{a} = \psi_{a}^{+} \gamma_0$ is the Dirac conjugated wave function with $\psi_{a}^{+}$ being its Hermitian conjugate, $\gamma^{\mu}_{A} = (\gamma_0, \bm{\gamma})$ are the Dirac matrices (indexes $A$ and $B$ refers to the corresponding atoms), $\varepsilon_n$ is the Dirac energy, and
\begin{eqnarray}
    \label{s2}
    S_{A}(x_1,x_2)=\frac{i}{2\pi}\int\limits_{-\infty}^{+\infty}d\Omega\, e^{-i\Omega(t_1-t_2)}
    \\\nonumber
    \times
    \sum\limits_{n}\frac{\psi^{A}_{n}(\bm{r}_1)\overline{\psi}^{A}_{n}(\bm{r}_2)}{\Omega - \varepsilon_{n}^{A} (1-i0)}
\end{eqnarray}
is the eigenstate decomposition of the electron propagator for atom $A$. The components of the photon propagator $D_{\mu\nu}$ can be expressed as the sum of two contributions: the zero-temperature part $D^{0}_{\mu\nu}$ and the thermal one $D^{\beta}_{\mu\nu}$, which accounts for the Planck frequency distribution associated with photons in the thermal reservoir \cite{SLP-QED}. The latter allows us to investigate and elucidate the influence of blackbody radiation on the interaction of two hydrogen atoms separated by a distance $R$.

To explore the effects resulting from the incorporation of distribution function, we turn to the finite-temperature Quantum Electrodynamics (TQED) approach formulated by Donoghue and Holstein (DH) in~\cite{DonH}. Then, the interaction between a free-electron gas (in the absence of an external field) and a photon gas is considered under thermal equilibrium conditions. This interaction is described using a grand canonical statistical operator, which alters both the electron and photon propagators. Our objective is to investigate the impact of blackbody radiation (BBR) on atomic levels. Therefore, we retain the electron propagator in the standard (zero-temperature) form (its thermal part is exponentially suppressed over temperature) and employ QED perturbation theory to account for the influence of BBR. This involves considering the thermal photon propagator only.

According to~\cite{DonH}, the photon propagator in the Feynman (F) gauge and momentum space reads as
\begin{eqnarray}
    \label{1}
    D_{\mu\nu}^{\mathrm{DH}}(k)=D^{0, F}_{\mu\nu} +D^{\beta, F}_{\mu\nu} 
    \\\nonumber 
    =
    -4\pi g_{\mu\nu}
    \left[\frac{i}{k^2+i0} + 2\pi \delta (k^2)n_{\beta}(|\bm{k}|) \right]
    \,,
\end{eqnarray}
where $g_{\mu\nu}$ is the metric tensor of Minkowski space, $k=(k_{0},\bm{k})$ is the four-dimensional momentum, $k^2 = k_{0}^2 - \bm{k}^2$,
and $n_{\beta} $ is defined as follows
\begin{eqnarray}
    \label{nbeta}
    n_{\beta}(x)=\frac{1}{\exp{(\frac{x}{k_{B}T})}-1}.
\end{eqnarray}
Here $k_{B}$ is the Boltzmann constant in relativistic units and $T$ is the temperature in Kelvin. Note that Eq.~(\ref{1}) differs from the corresponding expression (4) in~\cite{DonH} by the factor of $4\pi$, which reflects the definition of the charge units $e^2 = \alpha$ used in this paper. %and also in the textbook~\cite{Berest}).
In the coordinate space the photon propagator can be found by 4-dimensional Fourier transform
\begin{eqnarray}
    \label{2}
    D_{\mu\nu}^{\mathrm{DH}}(x_{1},x_{2})= -4\pi g_{\mu\nu}\int\frac{d^4k}{(2\pi)^4}e^{-ik(x_1-x_2)}    
    \\\nonumber
    \times
    \left[\frac{i}{k^2+i0} + 2\pi \delta (k^2)n_{\beta}(|\bm{k}|) \right]\,.
    \\\nonumber
\end{eqnarray}

In the nonrelativistic problem we are considering, it is convenient to use the  the temporal gauge (also known as the Weyl gauge)~\cite{Berest}. %, in which $\phi=0$~\cite{Berest}.
Then the components of zero-temperature part of photon propagator are:
\begin{eqnarray}
    D^{0}_{00}(k) = D^{0}_{0i}(k) = 0
\end{eqnarray}
\begin{eqnarray}
    D^{0}_{ij}(k)=\frac{4\pi i}{k^2}\left(\delta_{ij}-\frac{k_{i}k_{j}}{k_{0}^2} \right)
\end{eqnarray}
Similarly, for the  finite temperature part we have~\cite{Escobedo2008, Escobedo2010}:
\begin{eqnarray}
    D^{\beta}_{00}(k) = D^{\beta}_{0i}(k) = 0\,,
\end{eqnarray}
\begin{eqnarray}
    D^{\beta}_{ij}(k)= 8\pi^2 \delta(k^2)\left(\delta_{ij}-\frac{k_{i}k_{j}}{k_{0}^2} \right)n_{\beta}(|\bm{k}|)\,,
\end{eqnarray}
where $i,\,j=1,\,2,\,3$. The evaluation of corresponding coordinate representation of zero-temperature and thermal propagators in temporal gauge is considered in Appendix~\ref{appendix:A} with the final result given by equation
\begin{eqnarray}
    \label{D0final_main_text}
    D^{0,\beta}_{ij}(x_1,x_2)=
    \frac{i}{2\pi} \int\limits_{-\infty}^{\infty} dk_{0}e^{-ik_{0}(t_1-t_2)}
    %\\\nonumber
    %\times
    D^{0,\beta}_{ij}(k_{0},R),\qquad
\end{eqnarray}
where $R=|\bm{r}_{1}-\bm{r}_{2}|$ is the interatomic distance and
\begin{eqnarray}
    \label{DOij_main_text}
    D^{0}_{ij}(k_{0},R)
    =
    \left(\delta_{ij}+ \frac{\nabla_i \nabla_j}{k_0^2} 
    \right)
    \left\lbrace    -
    \frac{e^{i|k_0|R}}{R} \right\rbrace
    \,,
\end{eqnarray}
\begin{eqnarray}
    \label{Dbeta_ij_main_text}
    D^{\beta}_{ij}(k_{0},R)
    =
    \left(\delta_{ij}+ \frac{\nabla_i \nabla_j}{k_0^2} 
    \right)
    \\\nonumber
    \times
    \left\lbrace
    -\frac{e^{i|k_{0}|R}}{R} 
%     \right.
%         \\\nonumber
%     \left.
    + \frac{e^{-i|k_{0}|R}}{R}
    \right\rbrace
    n_{\beta}(|k_{0}|).\qquad
\end{eqnarray}

% Equation (\ref{2}) can be computed using the conventional method. Specifically, we can select the $z$-axis in the $k$ space to align with the direction of the vector $\bm{r}$. 

% \begin{eqnarray}
%     \label{s4}
%     I_{ij}(k_{0},R)=\left( 
%     \frac{\delta_{ij}}{R}
%     +\frac{\nabla^{A}_{i}\nabla^{B}_{j}}{k_{0}^2}
%     \right)
%     \left(-
%     \frac{e^{ik_{0} R}}{R}
%     \right)
% \end{eqnarray}
Subsisting Eq.~(\ref{D0final_main_text}) into Eq.~(\ref{s1}) and performing integration over the time variables, we find
\begin{eqnarray}
    \label{s5}
    S_{AB}^{(4), \,L} = -2\pi i \delta(\varepsilon_{a'}^{A}-\varepsilon_{a}^{A}+\varepsilon_{b'}^{B}-\varepsilon_{b}^{B}) U_{AB}^{(4), \,L}(R)
    \,.
\end{eqnarray}
Here the amplitude of the process can be expressed as
\begin{eqnarray}
    \label{s6}
    U_{AB}^{(4), \,L}(R)=i\int\limits_{-\infty}^{+\infty}\frac{d k_{0}}{2\pi}D_{il}( k_{0},R)D_{km}( k_{0},R)
    \\\nonumber
    \times
    \sum\limits_{nn'}V_{a'nna}^{(-)ik}( k_{0})V_{b'n'n'b}^{(+)lm}( k_{0})
    \,,
\end{eqnarray}
where
\begin{eqnarray}
    \label{s7}
   V_{a'nna}^{(-)ik}( k_{0}) =  \frac{\langle \psi^A_{a'}| \bm{\alpha}_{i} | \psi^A_{n}\rangle \langle \psi^A_{n}| \bm{\alpha}_{k} | \psi^A_{a}\rangle}{\varepsilon_{a'}^{A}- k_{0}-\varepsilon_{n}^{A}(1-i0)}
   \,,
\end{eqnarray}
\begin{eqnarray}
    V_{b'n'n'b}^{(+)lm}( k_{0}) 
    % \frac{\langle \psi_{b'}| \bm{\alpha}_{l} | \psi_{n'}\rangle \langle \psi_{n'}| \bm{\alpha}_{m} | \psi_{b}\rangle}{\varepsilon_{b}^{B}-\varepsilon_{a'}^{A}+\varepsilon_{a}^{A}+ k_{0}-\varepsilon_{n'}(1-i0)}
    % \\\nonumber
    =
    \frac{\langle \psi^B_{b'}| \bm{\alpha}_{l} | \psi^B_{n'}\rangle \langle \psi^B_{n'}| \bm{\alpha}_{m} | \psi^B_{b}\rangle}{\varepsilon_{b'}^{B}+ k_{0}-\varepsilon_{n'}^{B}(1-i0)}
    \,.
\end{eqnarray}
Each factor $D_{ij}(k_{0},R)=D_{ij}^{0}(k_{0},R) + D_{ij}^{\beta}(k_{0},R)$ is the sum of zero and finite temperature contributions to the photon propagator, specified in  the 'mixed' representation (i.e. in frequency-space domain, see~\cite{Berest}) by Eq.~(\ref{DOij_main_text}) and (\ref{Dbeta_ij_main_text}), respectively.

Similarly, for the crossed-ladder (CL) diagram given by Fig.~\ref{fig:1} (b) the corresponding $S$-matrix elements is
\begin{eqnarray}
    \label{s8}
    S_{AB}^{(4), \,CL}=(-ie)^4\int dx_1 dx_2 dx_3 dx_4\overline{\psi}_{a'}(x_1)\overline{\psi}_{b'}(x_3)\quad
    \\\nonumber
    \times
    \gamma^{\mu_1}_{A}D_{\mu_1\mu_4}(x_1,x_4)\gamma^{\mu_3}_{B}
    D_{\mu_2\mu_3}(x_2,x_3)\gamma^{\mu_4}_{B}
    \\\nonumber
    \times
   S_{A}(x_1,x_2)\gamma^{\mu_2}_{A} S_{B}(x_3,x_4)\psi_{a}(x_2)\psi_{b}(x_4)
    \,.
\end{eqnarray}
Integration over the time variables in Eq.~(\ref{s8}) yields
\begin{eqnarray}
    \label{s9}
    S_{AB}^{(4), \,CL} = -2\pi i \delta(\varepsilon_{a'}^{A}-\varepsilon_{a}^{A}+\varepsilon_{b'}^{B}-\varepsilon_{b}^{B}) U_{AB}^{(4), \,L}(R)
    ,\quad
\end{eqnarray}
where
\begin{eqnarray}
    \label{s10}
    U_{AB}^{(4), \,CL}(R)=i\int\limits_{-\infty}^{+\infty}\frac{d k_{0}}{2\pi}D_{il}( k_{0},R)D_{km}( k_{0},R)
    \\\nonumber
    \times
    \sum\limits_{nn'}V_{a'nna}^{(-)ik}( k_{0})V_{b'n'n'b}^{(-)lm}( k_{0})
    \,.
\end{eqnarray}
% and 
% \begin{eqnarray}
%     \label{s11}
%     \widetilde{V}_{b'n'n'b}^{lm}( k_{0})=\frac{\langle \psi_{b'}| \bm{\alpha}_{i} | \psi_{n'}\rangle \langle \psi_{n'}| \bm{\alpha}_{k} | \psi_{b}\rangle}{\varepsilon_{b}^{B}- k_{0}-\varepsilon_{n'}(1-i0)}
% \end{eqnarray}

For each of these two basic Feynman diagrams we need to consider three more contributions accounting the total symmetry of system of two nonequivalent atoms with exchanged indexes in Fig.~\ref{fig:1} as follows: 1) $a'\leftrightarrow b'$; 2) $a\leftrightarrow b$; and 3) simultaneous replacement of both states $a'\leftrightarrow b'$ and $a\leftrightarrow b$. Moreover, to all these four contributions, it is also necessary to add exactly the same set but with permuted $ k_{0}$ and $ k_{0}'$ (this permutation just leads to additional terms given by Eqs.~(\ref{s6}) and (\ref{s10}) in which $ k_{0}$ is just replaced by $- k_{0}$). Thus the total number of terms is sixteen. For further consideration of the long-range potential, it is convenient to assume that the initial and final states of both atoms of the same type $A$ and $B$ remains unchanged, i.e. we set $a'=a$ and $b'=b$.  The latter circumstance decreases the number of nonequivalent contributions for the case when one atom is in the excited state to eight. 

Collecting all terms together and going to the nonrelativistic limit with Foldy–Wouthuysen transformation which in the leading order implies $\psi^{+}\bm{\alpha}\psi \approx \phi^{+}\frac{\hat{\bm{p}}}{m}\phi$, where $\phi$ is the solution of Schr\"odinger equation and $\hat{\bm{p}}$ is the electron momentum operator, we find for the total interaction amplitude the following equation in the velocity form:
\begin{gather}
    \label{total}
    U^{(4),\,\mathrm{tot}}_{AB}(R)= i\int\limits_{-\infty}^{+\infty}\frac{dk_{0}}{2\pi} 
    D_{il}(k_{0},R)D_{km}(k_{0},R)\sum\limits_{nn'}
    \\\nonumber
    \times
    \left\lbrace
    \left[
    \frac{\langle \phi_{a}^A| \bm{p}_{i} | \phi_{n}^A\rangle \langle \phi_{n}^A| \bm{p}_{k} | \phi_{a}^A\rangle}{E_{a}^A-k_{0}-E_{n}^A(1-i0)}
    + 
    \frac{\langle \phi_{a}^A| \bm{p}_{i} | \phi_{n}^A\rangle \langle \phi_{n}^A| \bm{p}_{k} | \phi_{a}^A\rangle}{E_{a}^A+k_{0}-E_{n}^A(1-i0)}
    \right]
    \right.
    \\\nonumber
    \times\left[
    \frac{\langle \phi_{b}^B| \bm{p}_{l} | \phi_{n'}^B\rangle \langle \phi_{n'}^B| \bm{p}_{m} | \phi_{b}^B\rangle}{E_{b}^B-k_{0}-E_{n'}^B(1-i0)}
    + 
    \frac{\langle \phi_{b}^B| \bm{p}_{l} | \phi_{n'}^B\rangle \langle \phi_{n'}^B| \bm{p}_{m} | \phi_{b}^B\rangle}{E_{b}^B+k_{0}-E_{n'}^B(1-i0)}
    \right]
    \\\nonumber
    \pm
    \left[
    \frac{\langle \phi_{a}^A| \bm{p}_{i} | \phi_{n}^A\rangle \langle \phi_{n}^A| \bm{p}_{k} | \phi_{b}^A\rangle}{E_{a}^A-k_{0}-E_{n}^A(1-i0)}
    + 
    \frac{\langle \phi_{a}^A| \bm{p}_{i} | \phi_{n}^A\rangle \langle \phi_{n}^A| \bm{p}_{k} | \phi_{b}^A\rangle}{E_{a}^A+k_{0}-E_{n}^A(1-i0)}
    \right]
    \\\nonumber
    \times\left[
    \frac{\langle \phi_{a}^B| \bm{p}_{l} | \phi_{n'}^B\rangle \langle \phi_{n'}^B| \bm{p}_{m} | \phi_{b}^B\rangle}{E_{b}^B-k_{0}-E_{n'}^B(1-i0)}
    + 
    \frac{\langle \phi_{a}^B| \bm{p}_{l} | \phi_{n'}^B\rangle \langle \phi_{n'}^B| \bm{p}_{m} | \phi_{b}^B\rangle}{E_{b}^B+k_{0}-E_{n'}^B(1-i0)}
    \right]
    \,.
\end{gather}
Here $E_{n}$ in contrast to $\varepsilon_{n}$ is the eigenvalue related to the atomic state $n$ of nonrelativistic hamiltonian $H_{S}$. 
Note, that the second contribution in curly brackets comes with $\pm$ sign and has off-diagonal matrix elements in numerator. This problem was covered in detail in a series of works~\cite{Adhikari:2017, Adhikari:2017:2}, where it was shown that for the identical atoms in different states, the potential becomes dependent on the symmetry of the wave function of the diatomic system. In our approach, these results are automatically restored.

To reduce the integrand in the expression (\ref{total}) to the product of atomic polarizabilities it is useful to transfer to the length-form of the matrix elements in the above equation. This can be done via the well-known commutation relation  $ p_{i}=i[H_{S},r_{i}] $ and some algebra. It is important to note that in the nonrelativistic expression (\ref{total}), the contribution arising from summation over the negative spectrum in the initial fully-relativistic Eqs.~(\ref{s6}) and (\ref{s10}) has already been omitted for the sake of brevity but it is exactly canceled when passing to the length form using the mentioned commutation relation. A detailed discussion of the corresponding transformations can be found in the textbook \cite{Akhiezer} (see chapter 27 therein) and \cite{Jentschura_gauge1, Jentschura_gauge2, Jentschura_gauge3}. Then the interaction potential can be conveniently written in the length-form in terms of atomic polarizability tensors $\alpha_{ij}$ as follows
\begin{eqnarray}
\label{pretotal}
     U^{(4),\,\mathrm{tot}}_{AB}(R)=\frac{i}{2\pi}\int\limits_{-\infty}^{+\infty}dk_{0} k_{0}^4
    D_{il}(k_{0},R)D_{km}(k_{0},R)
    \\\nonumber
    \times
    \left[
    \alpha^{A}_{ik}(k_0{}) \alpha^{B}_{lm}(k_{0})
    \pm 
    \alpha^{\overline{A}B}_{ik}(k_0{}) \alpha^{A\overline{B}}_{lm}(k_{0})
    \right]
    \, ,
\end{eqnarray}
where, according to~\cite{Adhikari:2017, Adhikari:2017:2}, the notations for diagonal $\alpha^{A}_{ik}$ and off-diagonal $\alpha^{AB}_{ik}$ contributions are introduced. For atoms in $s$-states the both tensors can be written in terms of scalar polarizabilities 
\begin{eqnarray}
\label{A0}
    \alpha^{A}_{ik}=\delta_{ik}\alpha_{A}\, ,
\end{eqnarray}
where
\begin{eqnarray}
 \label{A-1}
    \alpha_{A}(k_{0}) = \frac{e^2}{3}\sum\limits_{\pm}\sum\limits_{n}\frac{\langle \phi_{a} | \bm{r}|\phi_{n}\rangle \langle \phi_{n}|\bm{r} | \phi_{a} \rangle}{E_{n}(1-i0)-E_{a}\pm k_{0}}\,,   
\end{eqnarray}
\begin{eqnarray}
 \label{A2}
    \alpha_{\overline{A}B}(k_{0}) = \frac{e^2}{3}\sum\limits_{\pm}\sum\limits_{n}\frac{\langle \phi_{a} | \bm{r}|\phi_{n}\rangle \langle \phi_{n}|\bm{r} | \phi_{b} \rangle}{E_{n}(1-i0)-E_{a}\pm k_{0}}\,,   
\end{eqnarray}
\begin{eqnarray}
 \label{A3}
    \alpha_{A\overline{B}}(k_{0}) = \frac{e^2}{3}\sum\limits_{\pm}\sum\limits_{n}\frac{\langle \phi_{a} | \bm{r}|\phi_{n}\rangle \langle \phi_{n}|\bm{r} | \phi_{b} \rangle}{E_{n}(1-i0)-E_{b}\pm k_{0}}\,.   
\end{eqnarray}

For further calculations, it is necessary to calculate the explicit form of the functions $D_{il}(k_{0},R)$ and $ D_{km}(k_{0},R)$ given by the sum of zero, Eq.~(\ref{DOij_main_text}), and finite temperature, Eq.~(\ref{Dbeta_ij_main_text}), contributions. This can be performed by noting that 
\begin{eqnarray}
\label{+}
\left(\delta_{ij}+ \frac{\nabla_i \nabla_j}{k_0^2} 
    \right)\left(-
    \frac{e^{i|k_0|R}}{R} \right)  
    G_{0,\beta}\quad
    \\\nonumber
    = \left( \delta_{ik}\left(1+\frac{i}{|k_{0}|R}
    -\frac{1}{k_{0}^2R^2}
    \right)
    \right.
    \\\nonumber
    \left.
    +
    \frac{x_{i}x_{k}}{R^2}
    \left(
    \frac{3}{k_{0}^2R^2}
    -\frac{3i}{|k_{0}|R}
    -1
    \right)
    \right)
   % \\\nonumber
    %\times
    \left(-
    \frac{e^{i|k_{0}|R}}{R}
    \right)
    G_{0,\beta}
    \,,
\end{eqnarray}  
and
\begin{eqnarray}
\label{-}
\left(\delta_{ij}+ \frac{\nabla_i \nabla_j}{k_0^2} 
    \right)\left(+
    \frac{e^{-i|k_0|R}}{R} \right)    
    G_{0,\beta}\quad
    \\\nonumber
    = \left( \delta_{ik}\left(1-\frac{i}{|k_{0}|R}
    -\frac{1}{k_{0}^2R^2}
    \right)
    \right.
    \\\nonumber
    \left.
    +
    \frac{x_{i}x_{k}}{R^2}
    \left(
    \frac{3}{k_{0}^2R^2}
    +\frac{3i}{|k_{0}|R}
    -1
    \right)
    \right)
    %\\\nonumber
    %\times
    \left(+
    \frac{e^{-i|k_{0}|R}}{R}
    \right)
    G_{0,\beta}
    \,,
\end{eqnarray} 
where $G_{0}=1$ and $G_{\beta}=n_{\beta}(|k_{0}|)$. 

Substituting Eqs.~(\ref{A0}), (\ref{A-1}) and the zero-temperature part of $D_{il}(k_{0},R)$ given by Eq.~(\ref{DOij_main_text}) into Eq.~(\ref{pretotal}), and taking into account Eq.~(\ref{+}), we obtain the following well-known expression for the interaction energy of two identical atoms ($A=B$) in the ground state at $T=0$:
\begin{eqnarray}
    \label{7}
    U^{0}(R)=\frac{i}{2\pi R^2}\int\limits_{-\infty}^{\infty} dk_{0} k_{0}^4
    \alpha_{A}(k_0{}) \alpha_{A}(k_{0})
    \\\nonumber
    \times
    e^{2i|k_{0}|R}
    F_{1}(|k_{0}|,R),
\end{eqnarray}
where 
\begin{eqnarray}
    \label{F1}
    F_{1}(k_{0},R) = 
    1 + \frac{2i}{k_{0}R}
    -\frac{5}{(k_{0}R)^2}
    %\\\nonumber
    -\frac{6i}{(k_{0}R)^3}
    +\frac{3}{(k_{0}R)^4}.\quad\quad
\end{eqnarray}
When one of the atoms is in the excited state ($A\neq B$) it is necessary to make a substitution 
\begin{eqnarray}
\label{symmetry}
\alpha_{A} \alpha_{A} \rightarrow \alpha_{A}\alpha_{B} \pm 
\alpha_{\overline{A}B}\alpha_{A\overline{B}}.
\end{eqnarray}
This generalization was recently obtained for the long-range interaction of two hydrogen atoms in the $1s$ and $2s$ states in~\cite{Adhikari:2017}. 

Similar equation can be obtained for the finite temperature part by substituting Eq.~(\ref{Dbeta_ij_main_text}) and taking into account Eq.~(\ref{+}) as well as (\ref{-}). Finally, we find the thermal correction to the long-range interaction as
 \begin{eqnarray}
    \label{mainresult}
    U^{\beta}(R)=\frac{i}{\pi R^2}\int\limits_{-\infty}^{\infty} dk_{0} k_{0}^4\alpha_{A}(k_{0})\alpha_{A}(k_{0})\quad
    \\\nonumber
    \times
    e^{2i|k_{0}|R}
    F_{1}(|k_{0}|,R)
    n_{\beta}(|k_{0}|)
    \\\nonumber
    -
    \frac{i}{\pi R^2}\int\limits_{-\infty}^{\infty } dk_{0} k_{0}^4\alpha_{A}(k_{0})\alpha_{A}(k_{0})
    F_{2}(|k_{0}|,R)
    n_{\beta}(|k_{0}|)
    \,,
\end{eqnarray}
where $F_{1}$ is given by Eq.~(\ref{F1}) and $F_{2}$ is
\begin{eqnarray}
\label{F2}
F_{2}(k_{0},R)=
    1 + 
    \frac{1}{(k_{0}R)^2}
    +\frac{3}{(k_{0}R)^4}
    \,.
\end{eqnarray}
Again, when one of the atoms is in the excited state, it is necessary to make a substitution given by Eq.~(\ref{symmetry}). Below we consider various boundaries of the thermal potential defined by the equation~(\ref{mainresult}).

Note that the second term in Eq.~(\ref{mainresult}) arises due to the presence of additional term $\frac{e^{-i|k_0|R}}{R}$ in the equation for $ D_{ij}^{\beta} $, see Eq.~(\ref{Dbeta_ij_main_text}). In contrast to the previous considerations~\cite{Ninham:1998, Goedecke:1999, Passante:2007} the obtained finite temperature potential consist of these two contributions. It is interesting to note that the second term with the same power expansion as in Eq.~(\ref{F2}) was recently obtained in~\cite{Sherkunov2009} for long-range interaction between two atoms embedded in external electromagnetic field. The crucial point is that according to the expression (31), both summands comprise a divergence at $k_{0} = 0$ arising from the $3/(k_{0} R)^4$ term multiplyed by the singular at zero function $n_{\beta}(k_{0})$ in Eqs.~(\ref{F1}) and (\ref{F2}), and only the total expression is infrared finite. Furthermore, it's worth highlighting that the sum of two exponentials (see Eq.~(\ref{Dbeta_ij_main_text})), which arises specifically in the thermal case, yields the correct result when calculating the temperature dependent self-energy of a bound electron \cite{SLP-QED}. In the non-relativistic limit, the expression found in \cite{SLP-QED} exactly coincides with the formula for the thermal Stark shift as obtained within the framework of quantum mechanical perturbation theory \cite{Farley}.

\section{Short-range limit of interatomic interaction at finite temperature}
\label{section2}

Before proceeding to the analysis of the asymptotic behavior of the temperature correction to the long-range potential, we consider various limits for the leading non-thermal contribution. 

In the short-range limit ($a_{0}\ll R \ll \lambda_{0}$, where $a_{0}\sim 1/(m\alpha Z)$ is the Bohr radius and $\lambda_{0}\sim 1/(m(\alpha Z)^2)$ is the typical atomic wave-length in relativistic units) the expression~(\ref{7}) can be reduced to
\begin{eqnarray}
    \label{7a}
    U^{0}(R)=-\frac{C_{6}}{R^6}
    ,
\end{eqnarray}
where the coefficient $C_6$ is defined by
\begin{eqnarray}
\label{C6}
    C_{6}=-\frac{3i}{\pi}\int\limits_{0}^{\infty} dk_{0}\alpha_{A}(k_{0})\alpha_{A}(k_{0})
    \,.
\end{eqnarray}
The integration over $k_{0}$ in Eq.~(\ref{C6}) can be carried out both analytically (with the use of residue theorem) and numerically. The summation over the entire spectrum in Eqs.~(\ref{A-1})-(\ref{A3}) is commonly performed numerically using, for example, the B-splines approach for solutions of the Schr\"odinger equation ~\cite{DKB}. In the numerical results below, we also treat hydrogen atoms in the limit of infinite nucleus mass. Note, that the integral in Eq.~(\ref{C6}) is purely real despite the fact that polarizability in general case is a complex quantity (see proof of this statement in Appendix~\ref{appendix:B}). As a consequence, the imaginary part of the interaction potential appears only in the following orders of decomposition over powers of $k_{0}R$ in Eq.~(\ref{7}), see, e.g.~\cite{Adhikari:2017:2, Jentschura2023}.

%For two identical hydrogen atoms in $A=B=1s$ states we find the well-known result $C_{6}=6.499$ a.u.
For two identical hydrogen atoms in states $A=B=1s$ we arrive at the known result $C_{6}=6.499$ in atomic units (hereafter a.u.). For the $1s-2s$ interaction the resulting $C_{6}$ constant becomes symmetry dependent according to the substitution given by Eq.~(\ref{symmetry}). This leads to the dispersion constant $C_{6}=176.735 \pm 27.98 $ a.u., which is in perfect agreement with the results given in~\cite{Adhikari:2017, Adhikari:2017:2}. As shown in these works, in the case of quasi-degenerated states (e.g, levels $2s$ and $2p$ in hydrogen) it is also necessary to consider separately the interval $\lambda_{0}\ll R \ll \lambda_{L}$ with $\lambda_{L} = \frac{1}{m\alpha (\alpha Z)^4}$ (the wavelength of the Lamb shift in relativistic units), in which the general dependence $R^{-6}$ still remains. Then in this range $C_{6}=121.489 \pm 46.61 $~a.u.

Let us proceed to the analysis of various asymptotics of the finite-temperature contribution to the interaction potential given by expression (\ref{mainresult}). As was emphasized in \cite{Chiu:1985} in the thermal case, the interatomic distance parameter $R$ correlates with temperature $T$. Understanding this phenomenon is straightforward, as the main contribution to the integral in the expression is provided by the poles of the function $\alpha_{A}(k_{0})\alpha_{B}(k_{0})$, which is within the region bounded by the function $k_{0}^4n_{\beta}(k_0)$. The latter has a maximum at $k_{0}\sim k_{B}T$ and exponentially decreasing wings. Thus, in the thermal case, the asymptotic behavior of the potential is determined by the behavior of the oscillating exponential factor $e^{2ik_{0}R}$ in two different regions: 1) the short-range (SR) limit $a_ {0}\ll R\ll \frac{1}{k_{0}}$, which also implies the thermal condition $a_{0}\ll R\ll \frac{1}{k_{B}T}$, and 2) the long-range (LR) limit $R\gg \frac{1}{k_{0}}$ with temperatures satisfying the inequality $R \gg \frac{1}{k_{B}T}$.

%In the short-range limit one can set $k_{0}R\ll 1$ and the expansion of the exponential factor in the first term into a Taylor series in the vicinity of small argument up to terms of the order of $O((k_{0}R))^6)$ leads to
In the short-range limit we can set $k_{0}R\ll 1$. Then, decomposing the exponential factor into a Taylor series in the vicinity of a small argument up to terms of order $O( (k_{0}R)^6)$, we have
\begin{widetext}
\begin{gather}
    \label{8}
    U^{\beta}_{SR}(R) \approx  \frac{2i}{\pi R^2}\int\limits_{0}^{\infty} dk_{0} k_{0}^4\alpha_{A}\left(k_{0}\right)\alpha_{B}\left(k_{0}\right)
    \\\nonumber
    \times
    \left[
    \frac{4}{15} i (k_{0}R)^5
    + \frac{2}{3} (k_{0}R)^4
    -\frac{4}{3} i (k_{0}R)^3
    - 2 (k_{0}R)^2
    + 2 i (k_{0}R)
    + 1
    + O((k_{0}R)^6)
    \right]
    F_1(k_0,R)
n_{\beta}\left(k_{0}\right)
    \\\nonumber
    -
    \frac{2i}{\pi R^2}\int\limits_{0}^{\infty} 
    dk_{0}
    k_{0}^4\alpha_{A}\left(k_{0}\right)\alpha_{B}\left(k_{0}\right)
    F_2(k_0,R)
    n_{\beta}\left(k_{0}\right)
    \,.
    \end{gather}
Substitution of $F_{1}$, $F_{2}$ leads to
    \begin{gather}
    \label{substraction}
    U^{\beta}_{SR}(R) =
    \frac{2i}{\pi R^2}\int\limits_{0}^{\infty} 
    dk_{0}
    k_{0}^4
    \alpha_{A}\left(k_{0}\right)\alpha_{B}\left(k_{0}\right)
    \left[ 1 
    +
    \frac{1}{(k_{0}R)^2}
    +\frac{3}{(k_{0}R)^4} + 
    \frac{22 i }{15 } (k_{0}R) -\frac{16}{15}(k_{0}R)^2+ O((k_{0}R)^3))
    \right]    
    n_{\beta}\left(k_{0}\right)
    \\\nonumber
    -
    \frac{2i}{\pi R^2}\int\limits_{0}^{\infty}
    dk_{0}
    k_{0}^4 \alpha_{A}\left(k_{0}\right)\alpha_{B}\left(k_{0}\right)
    \left[
    1 + 
    \frac{1}{(k_{0}R)^2}
    +\frac{3}{(k_{0}R)^4} 
    \right]
    n_{\beta}\left(k_{0}\right)
    \\\nonumber
    =
    -\frac{44}{15\pi R}\int\limits_{0}^{\infty} 
    dk_{0}
    k_{0}^5
    \alpha_{A}\left(k_{0}\right)\alpha_{B}\left(k_{0}\right) 
    n_{\beta}\left(k_{0}\right)
    -\frac{32i}{15}\int\limits_{0}^{\infty}
    dk_{0}
    k_{0}^6 \alpha_{A}\left(k_{0}\right)\alpha_{B}\left(k_{0}\right) n_{\beta}(k_{0})
    \,.
\end{gather}
\end{widetext}
From Eq. (\ref{substraction}) it is clearly seen that the terms proportional to the $1 
    +
    \frac{1}{(k_{0}R)^2}
    +\frac{3}{(k_{0}R)^4}$ (see the second line in the equation) are cancel out in the total result. Thus, the final expression represents the Coulomb behaviour plus constant contribution. The latter arises from the second integral in the last line of Eq.~(\ref{8}) and depends only on the particular atomic polarizabilities. The resulting leading term confirms the conclusion reached earlier in~\cite{Chiu:1985}, but with the caveat that both contributions under consideration, as will be shown below, may be complex numbers. In addition, by going to atomic units in Eq.~(\ref{substraction}), we find that the temperature dependence is of order $\alpha^5 $, which is consistent with the result of \cite{Barton2011} (see Eq.~(C2) there).

The final expression for the short-range limit can be written in the compact form with the dispersion constant $C_{1}^{\beta}$ as follows:
\begin{eqnarray}
\label{srange0}
    U^{\beta}_{SR}(R)=-\frac{C_{1}^{\beta}}{R} + C_{0}^{\beta}
    ,
\end{eqnarray}
where 
\begin{eqnarray}
    \label{srange1}
    C_{1}^{\beta} = \frac{44}{15\pi} \int\limits_{0}^{\infty} 
    dk_{0}
    k_{0}^5 \alpha_{A}(k_{0})\alpha_{B}(k_{0}) n_{\beta}(k_{0})\,,
\end{eqnarray}

\begin{eqnarray}
    \label{srange2}
    C_{0}^{\beta} =  -\frac{32\,i}{15}\int\limits_{0}^{\infty}
    dk_{0}
    k_{0}^6 \alpha_{A}\left(k_{0}\right)\alpha_{B}\left(k_{0}\right) n_{\beta}(k_{0})
    \,.
\end{eqnarray}

It is important to note here that, in contrast to the zero temperature case, both constants of the leading order of short-range limit are complex due to the definitions (\ref{A-1})-(\ref{A3}). %This arises due to the fact that the product of polarizabilities in the integrand function is multiplied by a factor $k_{0}^N n_{\beta}(k_{0})$, where $N$ is the positive integer number.
A detailed analysis of the origin of the imaginary contribution to $C_{1}^{\beta}$ and $C_{0}^{\beta}$ is presented in Appendix \ref{appendix:B}. From a physical standpoint, this signifies the manifestation of line broadening. This effect admits the analogy with the examining the imaginary part of the thermal self-energy correction for the bound electron~\cite{Farley,Jentschura:BBR:2008,SLP-QED,ZAS:2020:twoloop}.

Additionally, one can consider asymptotic of Eq.~(\ref{srange0}) when $k_{B}T\ll m (\alpha Z)^2$ (i.e. the temperature is much less than the binding energy). For the two hydrogen atoms ($Z=1$) in the ground states $A=B=1s$ this inequality is valid up to temperatures $T \sim 10^4$ K. This implies that dynamic polarizability $\alpha_{1s}(k_{0})$ can be replaced by its static value $\alpha_{1s}(0)$ which is purely real. Performing integration over the frequency, we obtain
\begin{eqnarray}
\label{C1}
    C_{1}^{\beta}=-\frac{352\pi^5}{945}(k_{B}T)^6 \alpha^2_{1s}(0)\,,
\end{eqnarray}
\begin{eqnarray}
\label{C0}
    C_{0}^{\beta}= 1536\,i\,\zeta (7) (k_{B}T)^7\alpha^2_{1s}(0)\,,
\end{eqnarray}
where $\zeta(s)$ is the Riemann zeta function. At $T=300$~K one can find $C_{1}^{\beta}=-3.51\times 10^{-26}$ a.u. and $C_{0}^{\beta}=-i\, 3.3\times 10^{-30} $ a.u. These estimates lead to energy shift and level broadening on the order of $10^{-10}$ Hz and $10^{-14}$ Hz, respectively, which are negligible. If one or both atoms are in an excited state, this estimate is no longer applicable. In this case it is necessary to take into account the quasi-degenerate states available in the sum over the entire spectrum in the expression for the polarizabilities of atoms. In particular, one should consider states of opposite parity separated by a Lamb shift or a fine-structure interval, which are of order $m\alpha (\alpha Z)^4$ and $m(\alpha Z)^4$, respectively.

\section{Long-range limit of interatomic interaction at finite temperature}
\label{section3}

In the long-range (LR) limit, $ k_{0}R\gg 1$ (this also implies $R\gg 1/(k_{B}T)$), only the first terms in Eqs.~(\ref{F1}) and (\ref{F2}) remain important, i.e. we can set $F_{1}=F_{2}=1$. This leads to the expression:
% Then for the first integral in Eq.~(\ref{8}) only frequency values $k_{0}\ll m(\alpha Z)^2$ are of significance, while for $k_{0}\gg m(\alpha Z)^2$ the integral is suppressed by the rapidly oscillating exponent factor $e^{2ik_{0}R}$. 
\begin{eqnarray}
    \label{LR0}
    U^{\beta}_{LR}(R) 
    \approx
    \frac{2i}{\pi R^2}\int\limits_{0}^{\infty}
    dk_{0}
    k_{0}^4
    \alpha_{A}\left(k_{0}\right)\alpha_{B}\left(k_{0}\right)
    %\\\nonumber
    %\times
    e^{2ik_{0}R}
     n_{\beta}\left(k_{0}\right)\quad\quad
     \\\nonumber
    -
    \frac{2i}{\pi R^2}\int\limits_{0}^{\infty}
    dk_{0}
    k_{0}^4 \alpha_{A}\left(k_{0}\right)\alpha_{B}\left(k_{0}\right)
    n_{\beta}\left(k_{0}\right)\,.
   %  \\\nonumber
   % \approx
   %  -\frac{2 i \psi ^{(4)}\left(1-2 i R (k_{B} T) \right)(k_{B} T) ^5}{\pi R^2}
   %  \alpha_{A}\left(0\right)\alpha_{B}\left(0\right)
   %  \\\nonumber
   %  -
   %  \frac{2i}{\pi R^2}\int\limits_{0}^{\infty} k_{0}^4 \alpha_{A}\left(k_{0}\right)\alpha_{B}\left(k_{0}\right)
   %  n_{\beta}\left(k_{0}\right)
\end{eqnarray}

In Eq.~(\ref{LR0}), the second term obviously falls off slower (due to the absence of a highly oscillating exponent under the integral) with increasing $R$ than the first term. Therefore, in the long-range limit for the leading contribution, we find
\begin{eqnarray}
    \label{LR1}
    U^{\beta}_{LR}(R) 
    \approx
    \frac{B_2^{\beta}}{R^2}
    ,
\end{eqnarray}
where
\begin{eqnarray}
\label{longrange}
    B_2^{\beta}= -
    \frac{2i}{\pi}\int\limits_{0}^{\infty}
    dk_{0}
     k_{0}^4 
    \alpha_{A}\left(k_{0}\right)\alpha_{B}\left(k_{0}\right)
    n_{\beta}\left(k_{0}\right)\,.
\end{eqnarray}
In complete analogy with the reasoning in section~\ref{section2}, this contribution also yields a complex number (see Appendix \ref{appendix:B}). The energy shift in the LR limit defined by Eq.~(\ref{LR1}) corresponds to the $R^{-2}$ dependence. Numerical evaluation for two $1s-1s$ atoms at $T=300$ K gives $B_2^{\beta} = 7.75\times 10^{-43} - i\,7.04\times 10^{-22} $ a.u., which is also insignificant compared to the present level of experimental accuracy.

% Consider again the case of two hydrogen atoms in their ground states and temperature regime  $k_{B}T\ll m (\alpha Z)^2$. Then in second integral one can again use static approximation for the atomic polarizabilities. Performing remaining integration over $k_{0}$ we find
% \begin{eqnarray}
% \label{LR1}
%     U^{\beta}_{LR}(R) 
%     \approx
%     -\frac{2 i \psi ^{(4)}\left(1-2 i R (k_{B} T) \right)(k_{B} T) ^5}{\pi R^2}
%     \alpha_{A}\left(0\right)\alpha_{B}\left(0\right)
% \end{eqnarray}

% The first term in the Eq.~(\ref{LR0}) in the limit of high $R\gg 1/(k_{B}T)$ behaves as follows
% \begin{eqnarray}
%     \label{limit_1}
%     \lim\limits_{R \rightarrow \infty}\left(  
%     -\frac{2 i \psi ^{(4)}\left(1-2 i R (k_{B} T) \right)(k_{B} T) ^5}{\pi R^2}
%     \right) 
%     \\\nonumber
%     \times \alpha_{A}\left(0\right)\alpha_{B}\left(0\right)
%     = \frac{3i}{4\pi}\frac{k_{B}T}{R^6} \alpha_{A}\left(0\right)\alpha_{B}\left(0\right)
% \end{eqnarray}

% \begin{table}[]
%     \centering
%     \begin{tabular}{ c c }
%     \hline
%        State  & $\alpha_{ns}(0)$, a.u. \\
%        \hline
%        1s  &  $4.5$  \\
%        2s  &  $120$  \\
%        3s  &  $1012.5$  \\
%        6s  &  $51192$  \\
%        8s  &  $276480$  \\
%        12s &  $3.05856\times 10^6$  \\
%        20s &    \\
%     \hline
%     \end{tabular}
%     \caption{Static dipole polarizabilities $\alpha_{ns}$(0) of hydrogen $ns$-states. All values are in atomic units}
%     \label{tab:1}
% \end{table}

% 
\section{Results and discussion}
\label{final_results}

In this study, a rigorous quantum electrodynamics derivation of the long-range potential between two atoms placed in an equilibrium thermal radiation environment is presented. We explore two asymptotic behaviors of the resulting expression (short-range and long-range limits). The ultimate outcome for the temperature correction is expressed by Eq.~(\ref{mainresult}), differing from the result obtained through a phenomenological generalization of the known zero-temperature result \cite{Boyer:1975,Ninham:1998}. The latter approach results to a different behavior in the short-range limit, leading to a significant overestimation of the interaction induced by the thermal environment compared to the result presented in this work. 

To analyze the statement above, we briefly compare our results with the estimates given in \cite{Milonni:1996,Passante:2007}. In the short-range limit, the following expression, see Eq.~(10) in \cite{Passante:2007}, for the total potential (zero + finite temperature) was found:
\begin{eqnarray}
\label{other}
    \tilde{U}^{0+\beta}(R)=-\frac{3\pi}{R^6}\int\limits_{0}^{\infty}dk_{0}
    \alpha_{A}\left(k_{0}\right)\alpha_{B}\left(k_{0}\right)
    \\\nonumber
    \times
    \coth\left({\frac{k_{0}}{2k_{B}T}}\right)\sin\left( 2k_{0}R\right)
    \,.
\end{eqnarray}
Assuming that $k_{0}R\ll 1$ the leading thermal contribution of Eq.~(\ref{other}) can be estimated as 
\begin{eqnarray}
\label{other2}
        \tilde{U}^{\beta}(R)\approx -\frac{12\pi}{R^5}\int\limits_{0}^{\infty}dk_{0} k_{0}
    \alpha_{A}\left(k_{0}\right)\alpha_{B}\left(k_{0}\right)
    % \\\nonumber
    % \times
    n_{\beta}(k_0)
    \,,
\end{eqnarray}
where we took into account equality $\coth \left( \frac{x}{2}\right) = 1 + 2n_{\beta}(x)$. This equation is also infrared finite and can be evaluated numerically. For the two hydrogen atoms in their ground states and $T=300$ K,  we find $\tilde{U}^{\beta}(R) = -\frac{8.38\times 10^{-7}}{R^5} $ a.u. At $R=10$ a.u., this leads to a thermal shift of the order of $8.38\times 10^{-12}$ a.u., which is much larger than the same shift defined by the Eq.~(\ref{srange0})
%$\frac{C_{1} ^{\beta}}{R}$
with the calculated coefficient $C_{1}^{\beta}$ given in Table~\ref{tab:1}. Based on the fundamentals of the theory used in this study (thermal radiation is treated rigorously in the framework of QED at finite temperature rather than phenomenologically), we conclude that the previous results on the thermal interaction between two atoms (\ref{other}) are significantly overestimated.% Thus, we can conclude that the thermal contributions in the expression (\ref{other}) are significantly overestimated.

Numerical calculations presented in Tables~\ref{tab:1} and \ref{tab:2} show that only for highly excited states the corresponding energy shift barely reaches a value on the order of $1$ Hz (for interatomic distances on the order of ten Bohr radii) at room temperature. However, even this value is far beyond the accuracy achievable in modern experiments measuring the transition frequencies of involving Rydberg states.%However, this is beyond the precision achievable in modern experiments for measuring the frequencies of transitions between Rydberg states.

The approach developed in this work also permits the generalization of thermal corrections to the interaction of an atom with a wall, taking into account the interaction of multiple distributed atoms. For an ensemble of atoms, the resultant contribution should be notably greater. This is confirmed experimentally in the case of a Bose-Einstein condensate of $^{87}$Rb atoms located a few microns from a dielectric substrate~\cite{Cornell:2007}. Furthermore, there exists a discrepancy between theory and experiment, notably concerning the interaction of an atom with a graphene layer \cite{Klimchitskaya:2018}. It is important to emphasize that the result we obtained here involves the disparity between a rigorous QED derivation and the phenomenological approach. %This scenario closely resembles the one explored in the work by Sherkunov \cite{Sherkunov:2005}, where the QED approach also contradicts results obtained within the framework of linear response theory.

\begin{table}[hbtp]
\caption{The real and imaginary parts of the dispersion coefficient $C^{\beta}_{1}$ (see Eq.~(\ref{C1})) for two hydrogen atoms in states $a$ and $b$ (first column) at different temperatures $T$ in Kelvin. All values are given in atomic units. The imaginary part of $C^{\beta}_{1}$ for two atoms in ground states ($1s-1s$) is completely insignificant and therefore is not presented. }
    \begin{tabular}{c c c c}
    \hline
       States $a-b$ & T=300 & T=1000 & T=10$^4$ \\
       \hline \\
         &    $\mathrm{Re}\, C^{\beta}_{1}$   & & \\
       1s-1s  & $-3.51\times 10^{-26}$ & $-4.84\times 10^{-23}$ & $ -8.77\times 10^{-17}$\\
       2s-2s  & $-2.53\times 10^{-23}$ & $-4.08 \times 10^{-20}$ &  $5.72\times 10^{-15}$ \\
       3s-3s  & $-2.02\times 10^{-21}$ & $-1.37\times 10^{-18}$ & $-5.48\times 10^{-15}$  \\
       4s-4s  & $-8.04\times 10^{-20}$ & $-3.91\times 10^{-17}$ &  $-1.66\times 10^{-14}$ \\
%       6s-6s  &  &  &   \\
       8s-8s  & $2.11\times 10^{-17}$ & $-1.21\times 10^{-15}$ & $-1.98\times 10^{-13} $  \\
       % 12s-12s & $-1.07\times 10^{-16}$  & $-2.05\times 10^{-15}$ &  $-2.15\times 10^{-13} $ \\
       % 20s-20s & $-2.05\times 10^{-16}$ & $-2.34\times 10^{-15}$ &  $-2.30\times 10^{-13}$ \\
       &    $\mathrm{Im}\, C^{\beta}_{1}$   & & \\
       % 1s-1s  &  &  & \\
       2s-2s  & $-1.03\times 10^{-42}$ & $-8.39\times 10^{-23}$ & $4.42\times 10^{-15}$  \\
       3s-3s  & $-3.76\times 10^{-25}$ & $-5.00\times 10^{-18}$ & $1.56\times 10^{-14}$  \\
       4s-4s  & $-3.62\times 10^{-20}$ & $-1.37\times 10^{-18}$ & $1.42\times 10^{-14}$  \\
%       6s-6s  &  &  &   \\
       8s-8s  & $ -2.14\times 10^{-16}$ & $ -9.87\times 10^{-16}$ &  $ -3.06\times 10^{-15} $\\
       % 12s-12s & $-1.31\times 10^{-15}$   & $ - 7.33\times 10^{-15}$ & $-9.18\times 10^{-14}$  \\
       % 20s-20s &   $-9.06\times 10^{-16}$& $-3.34\times 10^{-15}$ & $-3.56 \times 10^{-14}$  \\
       \hline
    \end{tabular}
        \label{tab:1}
\end{table}

\begin{table}[hbtp]
\caption{The real and imaginary parts of dispersion coefficient $C^{\beta}_{0}$ (see Eq.~(\ref{C0})) for two hydrogen atoms in states $a$ an $b$ (first column) at different temperatures $T$ in Kelvin. All the values are in atomic units. The real part of $C^{\beta}_{0}$ for the two atoms in ground states ($1s-1s$) is completely negligible  and therefore is not presented.}
    \begin{tabular}{c c c c}
    \hline
       States $a-b$ & T=300 & T=1000 & T=10$^4$ \\
       \hline \\
         &    $\mathrm{Re}\, C^{\beta}_{0}$   & & \\
       % 1s-1s  & $5.18\times 10^{-51}$ & $8.00\times 10^{-47}$ & $1.88\times 10^{-19}$\\
       2s-2s  & $1.34\times 10^{-46}$ & $9.24\times 10^{-26}$ &  $1.24\times 10^{-17}$ \\
       3s-3s  & $1.46\times 10^{-28}$ & $1.62\times 10^{-21}$ & $-1.55\times 10^{-17}$  \\
       4s-4s  & $6.02\times 10^{-24} $ & $-4.14\times 10^{-21}$ & $-9.44\times 10^{-18} $  \\
%       6s-6s  &  &  &   \\
       8s-8s  & $7.52\times 10^{-21}$ & $-3.29\times 10^{-20}$ & $2.97\times 10^{-22}$  \\
       % 12s-12s & $ -1.07 \times 10^{-16} $ &  &   \\
       % 20s-20s &  &  &   \\
       &    $\mathrm{Im}\, C^{\beta}_{0}$   & & \\
       1s-1s  & $-3.31\times 10^{-30}$ & $-1.52\times 10^{-26}$ & $-2.88\times 10^{-19}$\\
       2s-2s  & $-2.40\times 10^{-27}$ & $-1.37\times 10^{-23}$ &  $5.35\times 10^{^-18} $ \\
       3s-3s  & $-1.99\times 10^{-25}$ & $2.34\times 10^{-22}$ & $-1.46\times 10^{-17}$  \\
       4s-4s  & $-7.60\times 10^{-24}$ & $7.62\times 10^{-21}$ &  $-2.16\times 10^{-17}$ \\
%       6s-6s  &  &  &   \\
       8s-8s  & $-8.38\times 10^{-22}$ & $-1.71\times 10^{-19}$ & $-1.54\times 10^{-16}$  \\
       % 12s-12s & $ - 1.31\times 10^{-15}    $ &  &   \\
       % 20s-20s &  &  &   \\
       \hline
    \end{tabular}
        \label{tab:2}
\end{table}

\section{Acknowledgements}
This work was supported by the foundation for the advancement of mathematics and theoretical physics "BASIS" (grant No.~23-1-3-31-1) and President grant MK-4796.2022.1.2.~Evaluation of long-range contribution (section~\ref{section3}) was supported by the Russian Science Foundation under grant No.~22-12-00043.

\appendix
\renewcommand{\theequation}{A\arabic{equation}}
\setcounter{equation}{0}

\section{Fourier transform of photon propagators}
\label{appendix:A}

Following~\cite{Berest} we define the 4-dimensional Fourier transform of function $f(k)$ as follows
\begin{eqnarray}
\label{A1}
    f(x)=\int\frac{d^4k}{(2\pi)^4}e^{-ik x}f(k)\,.
\end{eqnarray}
To find the corresponding coordinate representation of the total photon propagator, one should apply the above transformation to the sum of the zero- and finite-temperature parts given in momentum space:
\begin{eqnarray}
\label{sum}
    D_{\mu\nu}(k)=D^{0}_{\mu\nu}(k)+D^{\beta}_{\mu\nu}(k)\,,
\end{eqnarray}
Then, in the temporal gauge the corresponding evaluation for the first term in Eq.~(\ref{sum}) yields
\begin{eqnarray}
    D^{0}_{ij}(x_1,x_2)=\int\frac{d^4k}{(2\pi)^4}e^{-ik(x_1-x_2)}D^{0}_{ij}(k)=\qquad\qquad
    \\\nonumber
    -4\pi i\int\frac{dk_{0}}{2\pi}e^{-ik_{0}(t_1-t_2)}\int\frac{d^3k}{(2\pi)^3}
    %\\\nonumber
    %\times
    \frac{e^{i\bm{k}(\bm{r}_1-\bm{r}_2)}}{k_{0}^2-\bm{k}^2}\left(\delta_{ij}-\frac{k_{i}k_{j}}{k_{0}^2} \right)
    \\\nonumber
    =-\frac{i}{2\pi} \int\limits_{-\infty}^{\infty} dk_{0}e^{-ik_{0}(t_1-t_2)}\left(\delta_{ij}+ \frac{\nabla_i \nabla_j}{k_0^2} 
    \right)
    %\left\lbrace    -
    \frac{e^{i|k_0|r_{12}}}{r_{12}} %\right\rbrace
.\qquad
\end{eqnarray}
Finally, we can write %the result of Fourier transform as follows
\begin{eqnarray}
    \label{D0final}
    D^{0}_{ij}(x_1,x_2)=
    \frac{i}{2\pi} \int\limits_{-\infty}^{\infty} dk_{0}e^{-ik_{0}(t_1-t_2)}
    %\\\nonumber
    %\times
    D^{0}_{ij}(k_{0},r_{12}),\qquad
\end{eqnarray}
where we introduced notation: 
\begin{eqnarray}
    \label{DOij}
    D^{0}_{ij}(k_{0},r)
    =
    -\left(\delta_{ij}+ \frac{\nabla_i \nabla_j}{k_0^2} 
    \right)
    %\left\lbrace    -
    \frac{e^{i|k_0|r_{12}}}{r} %\right\rbrace
    .
\end{eqnarray}

The temperature dependent part in Eq.~(\ref{sum}) is
\begin{eqnarray}
\label{DPbeta_fourier}
    D^{\beta}_{ij}(x_1,x_2)=8\pi^2\int\frac{d^4k}{(2\pi)^4}e^{-ik(x_1-x_2)}D^{\beta,P}_{ij}(k)\qquad
    \\\nonumber
    = 8\pi^2
    \int\frac{dk_{0}}{2\pi}e^{-ik_{0}(t_1-t_2)}\int\frac{d^3k}{(2\pi)^3}e^{i\bm{k}(\bm{r}_1-\bm{r}_2)}\frac{n_{\beta}(|\bm{k}|)}{2|\bm{k}|}\qquad
    \\\nonumber
   \times
    %\frac{1}{2|\bm{k}|}
    (\delta(k_{0}-|\bm{k}|)+\delta(k_{0}+|\bm{k}|))
    %\\\nonumber
    %\times
    \left(\delta_{ij}-\frac{k_{i}k_{j}}{k_{0}^2} \right)=\qquad
    \\\nonumber
    %=
    %\frac{1}{\pi}
    \int\limits_{-\infty}^{+\infty}dk_{0}e^{-ik_{0}(t_1-t_2)}
    \left(\delta_{ij} + \frac{\nabla_{i}\nabla_{j}}{k_{0}^2} \right)
    %\\\nonumber
    %\times
    \frac{\sin(|k_{0}|r_{12})}{\pi r_{12}}n_{\beta}(|k_{0}|).
\end{eqnarray}
For our purposes it is convenient to rewrite Eq.~(\ref{DPbeta_fourier}) in terms of two contributions:
\begin{eqnarray}
    D^{\beta}_{ij}(x_1,x_2)=
    \frac{i}{2\pi}
    \int\limits_{-\infty}^{+\infty}dk_{0}e^{-ik_{0}(t_1-t_2)}\qquad
    \\\nonumber
    \times
    \left(\delta_{ij} + \frac{\nabla_{i}\nabla_{j}}{k_{0}^2} \right)
    \left\lbrace
    -\frac{e^{i|k_{0}|r_{12}}}{r_{12}} 
       + \frac{e^{-i|k_{0}|r_{12}}}{r_{12}}
    \right\rbrace
    n_{\beta}(|k_{0}|).
\end{eqnarray}
Similar to Eq.~(\ref{D0final}), the result of the Fourier transform of the thermal part can be written as follows:
\begin{eqnarray}
    \label{Dbeta_final}
    D^{\beta}_{ij}(x_1,x_2)=
    \frac{i}{2\pi} \int\limits_{-\infty}^{\infty} dk_{0}e^{-ik_{0}(t_1-t_2)}
    %\\\nonumber
    %\times
    D^{\beta}_{ij}(k_{0},r_{12}),\qquad
\end{eqnarray}
together with the notation
\begin{eqnarray}
    \label{Dbeta_ij}
    D^{\beta}_{ij}(k_{0},r)
    =
    \left(\delta_{ij}+ \frac{\nabla_i \nabla_j}{k_0^2} 
    \right)
    \\\nonumber
    \times
    \left\lbrace
    -\frac{e^{i|k_{0}|r_{12}}}{r_{12}} 
%     \right.
%         \\\nonumber
%     \left.
    + \frac{e^{-i|k_{0}|r_{12}}}{r_{12}}
    \right\rbrace
    n_{\beta}(|k_{0}|).\qquad
\end{eqnarray}
The sum of two equations (\ref{DOij}) and (\ref{Dbeta_ij}) (often called 'mixed' representation of photon propagator \cite{Berest})
\begin{eqnarray}
    \label{summixed}
    D_{ij}(k_{0},r) = D^{0}_{ij}(k_{0},r) + D^{\beta}_{ij}(k_{0},r)
\end{eqnarray}
 is then used in the evaluation of interaction potential given by Eq.~(\ref{pretotal}).
 
%\bigskip

\renewcommand{\theequation}{B\arabic{equation}}
\setcounter{equation}{0}

\section{Imaginary part of integrals with squared polarizability}
\label{appendix:B}

In this Appendix we describe how the imaginary part of the integrals with the squared atomic polarizability contributes to the dispersion coefficients in the zero and finite temperature cases. Although in the present work all integrations are performed completely numerically, such an analysis would not be superfluous.

For this purpose we consider the following model integral arising in the Eqs.~(\ref{srange1}), (\ref{srange2}) and (\ref{longrange}) for the dispersion coefficients:
\begin{eqnarray}
    \label{b1}
    J=\int_{0}^{\infty}\frac{f(k_{0})d k_{0}}{(a-k_{0}- i 0)^2},
\end{eqnarray}
where $a$ is the real positive number and $f(k_{0})$ is a function that is analytical on the real semi-axis and guarantees the convergence of the $J$. With the use of Dirac prescription
\begin{eqnarray}
    \label{b2}
    \frac{1}{a-k_{0}- i 0}= \mathrm{P} \frac{1}{a-k_{0}}+i\pi \delta (k_{0}-a),
\end{eqnarray}
%equation~(\ref{b1}) can be simplified to
($P$ stands for integral in the sense of the principal value) the equation~(\ref{b1}) can be simplified to
 \begin{eqnarray}
     \label{b3}
     J=\int_{0}^{\infty}\frac{f(k_{0})d k_{0}}{(a-k_{0}- i 0)^2}
     = -\frac{\partial }{\partial a}\int_{0}^{\infty}\frac{f(k_{0})d k_{0}}{a-k_{0}- i 0}\qquad
          \\\nonumber
          =
     -\frac{\partial }{\partial a}\int_{0}^{\infty}
     \left(
     \mathrm{P} \frac{1}{a-k_{0}}+i\pi \delta (k_{0}-a)
     \right)f(k_{0})
     dk_{0}\qquad
     \\\nonumber
     =
     -\frac{\partial }{\partial a} P \int_{0}^{\infty} 
     \frac{f(k_{0})d k_{0}}{a-k_{0}}
     -i\pi \frac{\partial }{\partial a} f(a).\qquad
 \end{eqnarray}
The first term in the last line of Eq.~(\ref{b3}) is purely real, while the second is imaginary (for the real $f$) and nonzero for $f(k_{0})\neq const$. Since for the finite temperature case $f(k_0)=k_{0}^N n_{\beta}(k_{0})$ (here $N>1$ is the integer number) the corresponding dispersion coefficient $C_{1}^{\beta}$ and $C_{0}^{\beta}$, see Eqs.~(\ref{srange1}) and (\ref{srange1}), becomes complex. The same holds for $B_{2}^{\beta}$ given by Eq.~(\ref{longrange}).

\bibliography{mybibfile}

\begin{thebibliography}{35}%
\makeatletter
\providecommand \@ifxundefined [1]{%
 \@ifx{#1\undefined}
}%
\providecommand \@ifnum [1]{%
 \ifnum #1\expandafter \@firstoftwo
 \else \expandafter \@secondoftwo
 \fi
}%
\providecommand \@ifx [1]{%
 \ifx #1\expandafter \@firstoftwo
 \else \expandafter \@secondoftwo
 \fi
}%
\providecommand \natexlab [1]{#1}%
\providecommand \enquote  [1]{``#1''}%
\providecommand \bibnamefont  [1]{#1}%
\providecommand \bibfnamefont [1]{#1}%
\providecommand \citenamefont [1]{#1}%
\providecommand \href@noop [0]{\@secondoftwo}%
\providecommand \href [0]{\begingroup \@sanitize@url \@href}%
\providecommand \@href[1]{\@@startlink{#1}\@@href}%
\providecommand \@@href[1]{\endgroup#1\@@endlink}%
\providecommand \@sanitize@url [0]{\catcode `\\12\catcode `\$12\catcode
  `\&12\catcode `\#12\catcode `\^12\catcode `\_12\catcode `\%12\relax}%
\providecommand \@@startlink[1]{}%
\providecommand \@@endlink[0]{}%
\providecommand \url  [0]{\begingroup\@sanitize@url \@url }%
\providecommand \@url [1]{\endgroup\@href {#1}{\urlprefix }}%
\providecommand \urlprefix  [0]{URL }%
\providecommand \Eprint [0]{\href }%
\providecommand \doibase [0]{http://dx.doi.org/}%
\providecommand \selectlanguage [0]{\@gobble}%
\providecommand \bibinfo  [0]{\@secondoftwo}%
\providecommand \bibfield  [0]{\@secondoftwo}%
\providecommand \translation [1]{[#1]}%
\providecommand \BibitemOpen [0]{}%
\providecommand \bibitemStop [0]{}%
\providecommand \bibitemNoStop [0]{.\EOS\space}%
\providecommand \EOS [0]{\spacefactor3000\relax}%
\providecommand \BibitemShut  [1]{\csname bibitem#1\endcsname}%
\let\auto@bib@innerbib\@empty
%</preamble>
\bibitem [{\citenamefont {Casimir}\ and\ \citenamefont
  {Polder}(1948)}]{CasimirPolder:1948}%
  \BibitemOpen
  \bibfield  {author} {\bibinfo {author} {\bibfnamefont {H.~B.~G.}\
  \bibnamefont {Casimir}}\ and\ \bibinfo {author} {\bibfnamefont
  {D.}~\bibnamefont {Polder}},\ }\href {\doibase 10.1103/PhysRev.73.360}
  {\bibfield  {journal} {\bibinfo  {journal} {Phys. Rev.}\ }\textbf {\bibinfo
  {volume} {73}},\ \bibinfo {pages} {360} (\bibinfo {year} {1948})}\BibitemShut
  {NoStop}%
\bibitem [{\citenamefont {Lifshitz}(1956)}]{Lifshitz}%
  \BibitemOpen
  \bibfield  {author} {\bibinfo {author} {\bibfnamefont {E.~M.}\ \bibnamefont
  {Lifshitz}},\ }\href {https://www.osti.gov/biblio/4359646} {\bibfield
  {journal} {\bibinfo  {journal} {Soviet Phys. JETP}\ }\textbf {\bibinfo
  {volume} {2}} (\bibinfo {year} {1956})}\BibitemShut {NoStop}%
\bibitem [{\citenamefont {Chiu}\ \emph {et~al.}(1985)\citenamefont {Chiu},
  \citenamefont {Dicus}, \citenamefont {Joseph},\ and\ \citenamefont
  {Teplitz}}]{Chiu:1985}%
  \BibitemOpen
  \bibfield  {author} {\bibinfo {author} {\bibfnamefont {C.~B.}\ \bibnamefont
  {Chiu}}, \bibinfo {author} {\bibfnamefont {D.~A.}\ \bibnamefont {Dicus}},
  \bibinfo {author} {\bibfnamefont {K.~B.}\ \bibnamefont {Joseph}}, \ and\
  \bibinfo {author} {\bibfnamefont {V.~L.}\ \bibnamefont {Teplitz}},\ }\href
  {\doibase 10.1103/PhysRevA.31.1458} {\bibfield  {journal} {\bibinfo
  {journal} {Phys. Rev. A}\ }\textbf {\bibinfo {volume} {31}},\ \bibinfo
  {pages} {1458} (\bibinfo {year} {1985})}\BibitemShut {NoStop}%
\bibitem [{\citenamefont {Gorza}\ and\ \citenamefont
  {Ducloy}(2006)}]{Gorza2006}%
  \BibitemOpen
  \bibfield  {author} {\bibinfo {author} {\bibfnamefont {M.-P.}\ \bibnamefont
  {Gorza}}\ and\ \bibinfo {author} {\bibfnamefont {M.}~\bibnamefont {Ducloy}},\
  }\href {\doibase 10.1140/epjd/e2006-00239-3} {\bibfield  {journal} {\bibinfo
  {journal} {Eur. Phys. J. D}\ }\textbf {\bibinfo {volume} {40}},\ \bibinfo
  {pages} {343} (\bibinfo {year} {2006})}\BibitemShut {NoStop}%
\bibitem [{\citenamefont {Fujii}\ \emph {et~al.}(2022)\citenamefont {Fujii},
  \citenamefont {Hongo},\ and\ \citenamefont {Enss}}]{Fujii:2022}%
  \BibitemOpen
  \bibfield  {author} {\bibinfo {author} {\bibfnamefont {K.}~\bibnamefont
  {Fujii}}, \bibinfo {author} {\bibfnamefont {M.}~\bibnamefont {Hongo}}, \ and\
  \bibinfo {author} {\bibfnamefont {T.}~\bibnamefont {Enss}},\ }\href {\doibase
  10.1103/PhysRevLett.129.233401} {\bibfield  {journal} {\bibinfo  {journal}
  {Phys. Rev. Lett.}\ }\textbf {\bibinfo {volume} {129}},\ \bibinfo {pages}
  {233401} (\bibinfo {year} {2022})}\BibitemShut {NoStop}%
\bibitem [{\citenamefont {Dzyaloshinskii}\ \emph {et~al.}(1961)\citenamefont
  {Dzyaloshinskii}, \citenamefont {Lifshitz},\ and\ \citenamefont
  {Pitaevskii}}]{Dzyaloshinskii1961}%
  \BibitemOpen
  \bibfield  {author} {\bibinfo {author} {\bibfnamefont {I.~E.}\ \bibnamefont
  {Dzyaloshinskii}}, \bibinfo {author} {\bibfnamefont {E.~M.}\ \bibnamefont
  {Lifshitz}}, \ and\ \bibinfo {author} {\bibfnamefont {L.~P.}\ \bibnamefont
  {Pitaevskii}},\ }\href {\doibase 10.1070/PU1961v004n02ABEH003330} {\bibfield
  {journal} {\bibinfo  {journal} {Sov. Phys. Usp.}\ }\textbf {\bibinfo {volume}
  {4}},\ \bibinfo {pages} {153} (\bibinfo {year} {1961})}\BibitemShut {NoStop}%
\bibitem [{\citenamefont {Berman}\ \emph {et~al.}(2014)\citenamefont {Berman},
  \citenamefont {Ford},\ and\ \citenamefont {Milonni}}]{Berman:2014}%
  \BibitemOpen
  \bibfield  {author} {\bibinfo {author} {\bibfnamefont {P.~R.}\ \bibnamefont
  {Berman}}, \bibinfo {author} {\bibfnamefont {G.~W.}\ \bibnamefont {Ford}}, \
  and\ \bibinfo {author} {\bibfnamefont {P.~W.}\ \bibnamefont {Milonni}},\
  }\href {\doibase 10.1103/PhysRevA.89.022127} {\bibfield  {journal} {\bibinfo
  {journal} {Phys. Rev. A}\ }\textbf {\bibinfo {volume} {89}},\ \bibinfo
  {pages} {022127} (\bibinfo {year} {2014})}\BibitemShut {NoStop}%
\bibitem [{\citenamefont {Safari}\ and\ \citenamefont
  {Karimpour}(2015)}]{Safari:2015}%
  \BibitemOpen
  \bibfield  {author} {\bibinfo {author} {\bibfnamefont {H.}~\bibnamefont
  {Safari}}\ and\ \bibinfo {author} {\bibfnamefont {M.~R.}\ \bibnamefont
  {Karimpour}},\ }\href {\doibase 10.1103/PhysRevLett.114.013201} {\bibfield
  {journal} {\bibinfo  {journal} {Phys. Rev. Lett.}\ }\textbf {\bibinfo
  {volume} {114}},\ \bibinfo {pages} {013201} (\bibinfo {year}
  {2015})}\BibitemShut {NoStop}%
\bibitem [{\citenamefont {Obrecht}\ \emph {et~al.}(2007)\citenamefont
  {Obrecht}, \citenamefont {Wild}, \citenamefont {Antezza}, \citenamefont
  {Pitaevskii}, \citenamefont {Stringari},\ and\ \citenamefont
  {Cornell}}]{Cornell:2007}%
  \BibitemOpen
  \bibfield  {author} {\bibinfo {author} {\bibfnamefont {J.~M.}\ \bibnamefont
  {Obrecht}}, \bibinfo {author} {\bibfnamefont {R.~J.}\ \bibnamefont {Wild}},
  \bibinfo {author} {\bibfnamefont {M.}~\bibnamefont {Antezza}}, \bibinfo
  {author} {\bibfnamefont {L.~P.}\ \bibnamefont {Pitaevskii}}, \bibinfo
  {author} {\bibfnamefont {S.}~\bibnamefont {Stringari}}, \ and\ \bibinfo
  {author} {\bibfnamefont {E.~A.}\ \bibnamefont {Cornell}},\ }\href {\doibase
  10.1103/PhysRevLett.98.063201} {\bibfield  {journal} {\bibinfo  {journal}
  {Phys. Rev. Lett.}\ }\textbf {\bibinfo {volume} {98}},\ \bibinfo {pages}
  {063201} (\bibinfo {year} {2007})}\BibitemShut {NoStop}%
\bibitem [{\citenamefont {Passerat~de Silans}\ \emph
  {et~al.}(2014)\citenamefont {Passerat~de Silans}, \citenamefont {Laliotis},
  \citenamefont {Maurin}, \citenamefont {Gorza}, \citenamefont {Chaves~de
  Souza~Segundo}, \citenamefont {Ducloy},\ and\ \citenamefont
  {Bloch}}]{PasseratdeSilans2014}%
  \BibitemOpen
  \bibfield  {author} {\bibinfo {author} {\bibfnamefont {T.}~\bibnamefont
  {Passerat~de Silans}}, \bibinfo {author} {\bibfnamefont {A.}~\bibnamefont
  {Laliotis}}, \bibinfo {author} {\bibfnamefont {I.}~\bibnamefont {Maurin}},
  \bibinfo {author} {\bibfnamefont {M.-P.}\ \bibnamefont {Gorza}}, \bibinfo
  {author} {\bibfnamefont {P.}~\bibnamefont {Chaves~de Souza~Segundo}},
  \bibinfo {author} {\bibfnamefont {M.}~\bibnamefont {Ducloy}}, \ and\ \bibinfo
  {author} {\bibfnamefont {D.}~\bibnamefont {Bloch}},\ }\href {\doibase
  10.1088/1054-660X/24/7/074009} {\bibfield  {journal} {\bibinfo  {journal}
  {Laser Phys.}\ }\textbf {\bibinfo {volume} {24}},\ \bibinfo {pages} {074009}
  (\bibinfo {year} {2014})}\BibitemShut {NoStop}%
\bibitem [{\citenamefont {Ninham}\ and\ \citenamefont
  {Daicic}(1998)}]{Ninham:1998}%
  \BibitemOpen
  \bibfield  {author} {\bibinfo {author} {\bibfnamefont {B.~W.}\ \bibnamefont
  {Ninham}}\ and\ \bibinfo {author} {\bibfnamefont {J.}~\bibnamefont
  {Daicic}},\ }\href {\doibase 10.1103/PhysRevA.57.1870} {\bibfield  {journal}
  {\bibinfo  {journal} {Phys. Rev. A}\ }\textbf {\bibinfo {volume} {57}},\
  \bibinfo {pages} {1870} (\bibinfo {year} {1998})}\BibitemShut {NoStop}%
\bibitem [{\citenamefont {Goedecke}\ and\ \citenamefont
  {Wood}(1999)}]{Goedecke:1999}%
  \BibitemOpen
  \bibfield  {author} {\bibinfo {author} {\bibfnamefont {G.~H.}\ \bibnamefont
  {Goedecke}}\ and\ \bibinfo {author} {\bibfnamefont {R.~C.}\ \bibnamefont
  {Wood}},\ }\href {\doibase 10.1103/PhysRevA.60.2577} {\bibfield  {journal}
  {\bibinfo  {journal} {Phys. Rev. A}\ }\textbf {\bibinfo {volume} {60}},\
  \bibinfo {pages} {2577} (\bibinfo {year} {1999})}\BibitemShut {NoStop}%
\bibitem [{\citenamefont {Passante}\ and\ \citenamefont
  {Spagnolo}(2007)}]{Passante:2007}%
  \BibitemOpen
  \bibfield  {author} {\bibinfo {author} {\bibfnamefont {R.}~\bibnamefont
  {Passante}}\ and\ \bibinfo {author} {\bibfnamefont {S.}~\bibnamefont
  {Spagnolo}},\ }\href {\doibase 10.1103/PhysRevA.76.042112} {\bibfield
  {journal} {\bibinfo  {journal} {Phys. Rev. A}\ }\textbf {\bibinfo {volume}
  {76}},\ \bibinfo {pages} {042112} (\bibinfo {year} {2007})}\BibitemShut
  {NoStop}%
\bibitem [{\citenamefont {Lindgren}(2011)}]{lindgren}%
  \BibitemOpen
  \bibfield  {author} {\bibinfo {author} {\bibfnamefont {I.}~\bibnamefont
  {Lindgren}},\ }\href {\doibase 10.1007/978-3-319-15386-5} {\emph {\bibinfo
  {title} {Relativistic Many-Body Theory}}},\ Vol.~\bibinfo {volume} {63}\
  (\bibinfo  {publisher} {Springer-Verlag New York},\ \bibinfo {year}
  {2011})\BibitemShut {NoStop}%
\bibitem [{\citenamefont {Solovyev}\ \emph {et~al.}(2015)\citenamefont
  {Solovyev}, \citenamefont {Labzowsky},\ and\ \citenamefont
  {Plunien}}]{SLP-QED}%
  \BibitemOpen
  \bibfield  {author} {\bibinfo {author} {\bibfnamefont {D.}~\bibnamefont
  {Solovyev}}, \bibinfo {author} {\bibfnamefont {L.}~\bibnamefont {Labzowsky}},
  \ and\ \bibinfo {author} {\bibfnamefont {G.}~\bibnamefont {Plunien}},\ }\href
  {\doibase 10.1103/PhysRevA.92.022508} {\bibfield  {journal} {\bibinfo
  {journal} {Phys. Rev. A}\ }\textbf {\bibinfo {volume} {92}},\ \bibinfo
  {pages} {022508} (\bibinfo {year} {2015})}\BibitemShut {NoStop}%
\bibitem [{\citenamefont {Donoghue}\ and\ \citenamefont
  {Holstein}(1983)}]{DonH}%
  \BibitemOpen
  \bibfield  {author} {\bibinfo {author} {\bibfnamefont {J.~F.}\ \bibnamefont
  {Donoghue}}\ and\ \bibinfo {author} {\bibfnamefont {B.~R.}\ \bibnamefont
  {Holstein}},\ }\href {\doibase 10.1103/PhysRevD.28.340 Erratum Phys. Rev.,
  D,29, 3004 (1984)} {\bibfield  {journal} {\bibinfo  {journal} {Phys. Rev. D}\
  }\textbf {\bibinfo {volume} {28(R)}},\ \bibinfo {pages} {3400} (\bibinfo
  {year} {1983})}\BibitemShut {NoStop}%
\bibitem [{\citenamefont {Berestetskii}\ \emph {et~al.}(1982)\citenamefont
  {Berestetskii}, \citenamefont {Lifshits},\ and\ \citenamefont
  {Pitaevskii}}]{Berest}%
  \BibitemOpen
  \bibfield  {author} {\bibinfo {author} {\bibfnamefont {V.}~\bibnamefont
  {Berestetskii}}, \bibinfo {author} {\bibfnamefont {E.}~\bibnamefont
  {Lifshits}}, \ and\ \bibinfo {author} {\bibfnamefont {L.}~\bibnamefont
  {Pitaevskii}},\ }\href@noop {} {\emph {\bibinfo {title} {Quantum
  Electrodynamics}}}\ (\bibinfo  {publisher} {Oxford Butterworth-Heinemann},\
  \bibinfo {year} {1982})\BibitemShut {NoStop}%
\bibitem [{\citenamefont {Escobedo}\ and\ \citenamefont
  {Soto}(2008)}]{Escobedo2008}%
  \BibitemOpen
  \bibfield  {author} {\bibinfo {author} {\bibfnamefont {M.~A.}\ \bibnamefont
  {Escobedo}}\ and\ \bibinfo {author} {\bibfnamefont {J.}~\bibnamefont
  {Soto}},\ }\href {\doibase 10.1103/PhysRevA.78.032520} {\bibfield  {journal}
  {\bibinfo  {journal} {Phys. Rev. A}\ }\textbf {\bibinfo {volume} {78}},\
  \bibinfo {pages} {437} (\bibinfo {year} {2008})}\BibitemShut {NoStop}%
\bibitem [{\citenamefont {Escobedo}\ and\ \citenamefont
  {Soto}(2010)}]{Escobedo2010}%
  \BibitemOpen
  \bibfield  {author} {\bibinfo {author} {\bibfnamefont {M.~A.}\ \bibnamefont
  {Escobedo}}\ and\ \bibinfo {author} {\bibfnamefont {J.}~\bibnamefont
  {Soto}},\ }\href {\doibase 10.1103/PhysRevA.82.042506} {\bibfield  {journal}
  {\bibinfo  {journal} {Phys. Rev. A}\ }\textbf {\bibinfo {volume} {82}}
  (\bibinfo {year} {2010}),\ 10.1103/PhysRevA.82.042506}\BibitemShut {NoStop}%
\bibitem [{\citenamefont {Adhikari}\ \emph
  {et~al.}(2017{\natexlab{a}})\citenamefont {Adhikari}, \citenamefont
  {Debierre}, \citenamefont {Matveev}, \citenamefont {Kolachevsky},\ and\
  \citenamefont {Jentschura}}]{Adhikari:2017}%
  \BibitemOpen
  \bibfield  {author} {\bibinfo {author} {\bibfnamefont {C.~M.}\ \bibnamefont
  {Adhikari}}, \bibinfo {author} {\bibfnamefont {V.}~\bibnamefont {Debierre}},
  \bibinfo {author} {\bibfnamefont {A.}~\bibnamefont {Matveev}}, \bibinfo
  {author} {\bibfnamefont {N.}~\bibnamefont {Kolachevsky}}, \ and\ \bibinfo
  {author} {\bibfnamefont {U.~D.}\ \bibnamefont {Jentschura}},\ }\href
  {\doibase 10.1103/PhysRevA.95.022703} {\bibfield  {journal} {\bibinfo
  {journal} {Phys. Rev. A}\ }\textbf {\bibinfo {volume} {95}},\ \bibinfo
  {pages} {022703} (\bibinfo {year} {2017}{\natexlab{a}})}\BibitemShut
  {NoStop}%
\bibitem [{\citenamefont {Adhikari}\ \emph
  {et~al.}(2017{\natexlab{b}})\citenamefont {Adhikari}, \citenamefont
  {Debierre},\ and\ \citenamefont {Jentschura}}]{Adhikari:2017:2}%
  \BibitemOpen
  \bibfield  {author} {\bibinfo {author} {\bibfnamefont {C.~M.}\ \bibnamefont
  {Adhikari}}, \bibinfo {author} {\bibfnamefont {V.}~\bibnamefont {Debierre}},
  \ and\ \bibinfo {author} {\bibfnamefont {U.~D.}\ \bibnamefont {Jentschura}},\
  }\href {\doibase 10.1103/PhysRevA.96.032702} {\bibfield  {journal} {\bibinfo
  {journal} {Phys. Rev. A}\ }\textbf {\bibinfo {volume} {96}},\ \bibinfo
  {pages} {032702} (\bibinfo {year} {2017}{\natexlab{b}})}\BibitemShut
  {NoStop}%
\bibitem [{\citenamefont {Akhiezer}\ and\ \citenamefont
  {Berestetskii}(1965)}]{Akhiezer}%
  \BibitemOpen
  \bibfield  {author} {\bibinfo {author} {\bibfnamefont {A.~I.}\ \bibnamefont
  {Akhiezer}}\ and\ \bibinfo {author} {\bibfnamefont {V.~B.}\ \bibnamefont
  {Berestetskii}},\ }\href@noop {} {\emph {\bibinfo {title} {Quantum
  Electrodynamics}}}\ (\bibinfo  {publisher} {Wiley-Interscience, New York},\
  \bibinfo {year} {1965})\BibitemShut {NoStop}%
\bibitem [{\citenamefont {Jentschura}(2016)}]{Jentschura_gauge1}%
  \BibitemOpen
  \bibfield  {author} {\bibinfo {author} {\bibfnamefont {U.~D.}\ \bibnamefont
  {Jentschura}},\ }\href {\doibase 10.1103/PhysRevA.94.022117} {\bibfield
  {journal} {\bibinfo  {journal} {Phys. Rev. A}\ }\textbf {\bibinfo {volume}
  {94}},\ \bibinfo {pages} {022117} (\bibinfo {year} {2016})}\BibitemShut
  {NoStop}%
\bibitem [{\citenamefont {Jentschura}\ and\ \citenamefont
  {Adhikari}(2018{\natexlab{a}})}]{Jentschura_gauge2}%
  \BibitemOpen
  \bibfield  {author} {\bibinfo {author} {\bibfnamefont {U.~D.}\ \bibnamefont
  {Jentschura}}\ and\ \bibinfo {author} {\bibfnamefont {C.~M.}\ \bibnamefont
  {Adhikari}},\ }\href {\doibase 10.1103/PhysRevA.97.062120} {\bibfield
  {journal} {\bibinfo  {journal} {Phys. Rev. A}\ }\textbf {\bibinfo {volume}
  {97}},\ \bibinfo {pages} {062120} (\bibinfo {year}
  {2018}{\natexlab{a}})}\BibitemShut {NoStop}%
\bibitem [{\citenamefont {Jentschura}\ and\ \citenamefont
  {Adhikari}(2018{\natexlab{b}})}]{Jentschura_gauge3}%
  \BibitemOpen
  \bibfield  {author} {\bibinfo {author} {\bibfnamefont {U.~D.}\ \bibnamefont
  {Jentschura}}\ and\ \bibinfo {author} {\bibfnamefont {C.~M.}\ \bibnamefont
  {Adhikari}},\ }\href {\doibase 10.1103/PhysRevA.97.062120} {\bibfield
  {journal} {\bibinfo  {journal} {Phys. Rev. A}\ }\textbf {\bibinfo {volume}
  {97}},\ \bibinfo {pages} {062120} (\bibinfo {year}
  {2018}{\natexlab{b}})}\BibitemShut {NoStop}%
\bibitem [{\citenamefont {Sherkunov}(2009)}]{Sherkunov2009}%
  \BibitemOpen
  \bibfield  {author} {\bibinfo {author} {\bibfnamefont {Y.}~\bibnamefont
  {Sherkunov}},\ }\href {\doibase 10.1088/1742-6596/161/1/012041} {\bibfield
  {journal} {\bibinfo  {journal} {J. Phys.: Conf. Ser.}\ }\textbf {\bibinfo
  {volume} {161}},\ \bibinfo {pages} {012041} (\bibinfo {year}
  {2009})}\BibitemShut {NoStop}%
\bibitem [{\citenamefont {Farley}\ and\ \citenamefont {Wing}(1981)}]{Farley}%
  \BibitemOpen
  \bibfield  {author} {\bibinfo {author} {\bibfnamefont {J.~W.}\ \bibnamefont
  {Farley}}\ and\ \bibinfo {author} {\bibfnamefont {W.~H.}\ \bibnamefont
  {Wing}},\ }\href {\doibase 10.1103/PhysRevA.23.2397} {\bibfield  {journal}
  {\bibinfo  {journal} {Phys. Rev. A}\ }\textbf {\bibinfo {volume} {23}},\
  \bibinfo {pages} {2397} (\bibinfo {year} {1981})}\BibitemShut {NoStop}%
\bibitem [{\citenamefont {Shabaev}\ \emph {et~al.}(2004)\citenamefont
  {Shabaev}, \citenamefont {Tupitsyn}, \citenamefont {Yerokhin}, \citenamefont
  {Plunien},\ and\ \citenamefont {Soff}}]{DKB}%
  \BibitemOpen
  \bibfield  {author} {\bibinfo {author} {\bibfnamefont {V.~M.}\ \bibnamefont
  {Shabaev}}, \bibinfo {author} {\bibfnamefont {I.~I.}\ \bibnamefont
  {Tupitsyn}}, \bibinfo {author} {\bibfnamefont {V.~A.}\ \bibnamefont
  {Yerokhin}}, \bibinfo {author} {\bibfnamefont {G.}~\bibnamefont {Plunien}}, \
  and\ \bibinfo {author} {\bibfnamefont {G.}~\bibnamefont {Soff}},\ }\href
  {\doibase 10.1103/PhysRevLett.93.130405} {\bibfield  {journal} {\bibinfo
  {journal} {Phys. Rev. Lett.}\ }\textbf {\bibinfo {volume} {93}},\ \bibinfo
  {pages} {130405} (\bibinfo {year} {2004})}\BibitemShut {NoStop}%
\bibitem [{\citenamefont {Jentschura}\ and\ \citenamefont
  {Adhikari}(2023)}]{Jentschura2023}%
  \BibitemOpen
  \bibfield  {author} {\bibinfo {author} {\bibfnamefont {U.~D.}\ \bibnamefont
  {Jentschura}}\ and\ \bibinfo {author} {\bibfnamefont {C.~M.}\ \bibnamefont
  {Adhikari}},\ }\href {\doibase 10.3390/atoms11010010} {\bibfield  {journal}
  {\bibinfo  {journal} {Atoms}\ }\textbf {\bibinfo {volume} {11}},\ \bibinfo
  {pages} {10} (\bibinfo {year} {2023})}\BibitemShut {NoStop}%
\bibitem [{\citenamefont {Barton}(2011)}]{Barton2011}%
  \BibitemOpen
  \bibfield  {author} {\bibinfo {author} {\bibfnamefont {G.}~\bibnamefont
  {Barton}},\ }\href {\doibase 10.1088/1367-2630/13/4/043023} {\bibfield
  {journal} {\bibinfo  {journal} {New J. Phys.}\ }\textbf {\bibinfo {volume}
  {13}},\ \bibinfo {pages} {043023} (\bibinfo {year} {2011})}\BibitemShut
  {NoStop}%
\bibitem [{\citenamefont {Jentschura}\ and\ \citenamefont
  {Haas}(2008)}]{Jentschura:BBR:2008}%
  \BibitemOpen
  \bibfield  {author} {\bibinfo {author} {\bibfnamefont {U.~D.}\ \bibnamefont
  {Jentschura}}\ and\ \bibinfo {author} {\bibfnamefont {M.}~\bibnamefont
  {Haas}},\ }\href {\doibase 10.1103/PhysRevA.78.042504} {\bibfield  {journal}
  {\bibinfo  {journal} {Phys. Rev. A}\ }\textbf {\bibinfo {volume} {78}},\
  \bibinfo {pages} {042504} (\bibinfo {year} {2008})}\BibitemShut {NoStop}%
\bibitem [{\citenamefont {Zalialiutdinov}\ \emph {et~al.}(2020)\citenamefont
  {Zalialiutdinov}, \citenamefont {Anikin},\ and\ \citenamefont
  {Solovyev}}]{ZAS:2020:twoloop}%
  \BibitemOpen
  \bibfield  {author} {\bibinfo {author} {\bibfnamefont {T.}~\bibnamefont
  {Zalialiutdinov}}, \bibinfo {author} {\bibfnamefont {A.}~\bibnamefont
  {Anikin}}, \ and\ \bibinfo {author} {\bibfnamefont {D.}~\bibnamefont
  {Solovyev}},\ }\href {\doibase 10.1103/PhysRevA.102.032204} {\bibfield
  {journal} {\bibinfo  {journal} {Phys. Rev. A}\ }\textbf {\bibinfo {volume}
  {102}},\ \bibinfo {pages} {032204} (\bibinfo {year} {2020})}\BibitemShut
  {NoStop}%
\bibitem [{\citenamefont {Boyer}(1975)}]{Boyer:1975}%
  \BibitemOpen
  \bibfield  {author} {\bibinfo {author} {\bibfnamefont {T.~H.}\ \bibnamefont
  {Boyer}},\ }\href {\doibase 10.1103/PhysRevA.11.1650} {\bibfield  {journal}
  {\bibinfo  {journal} {Phys. Rev. A}\ }\textbf {\bibinfo {volume} {11}},\
  \bibinfo {pages} {1650} (\bibinfo {year} {1975})}\BibitemShut {NoStop}%
\bibitem [{\citenamefont {Milonni}\ and\ \citenamefont
  {Smith}(1996)}]{Milonni:1996}%
  \BibitemOpen
  \bibfield  {author} {\bibinfo {author} {\bibfnamefont {P.~W.}\ \bibnamefont
  {Milonni}}\ and\ \bibinfo {author} {\bibfnamefont {A.}~\bibnamefont
  {Smith}},\ }\href {\doibase 10.1103/PhysRevA.53.3484} {\bibfield  {journal}
  {\bibinfo  {journal} {Phys. Rev. A}\ }\textbf {\bibinfo {volume} {53}},\
  \bibinfo {pages} {3484} (\bibinfo {year} {1996})}\BibitemShut {NoStop}%
\bibitem [{\citenamefont {Klimchitskaya}\ and\ \citenamefont
  {Mostepanenko}(2018)}]{Klimchitskaya:2018}%
  \BibitemOpen
  \bibfield  {author} {\bibinfo {author} {\bibfnamefont {G.~L.}\ \bibnamefont
  {Klimchitskaya}}\ and\ \bibinfo {author} {\bibfnamefont {V.~M.}\ \bibnamefont
  {Mostepanenko}},\ }\href {\doibase 10.1103/PhysRevA.98.032506} {\bibfield
  {journal} {\bibinfo  {journal} {Phys. Rev. A}\ }\textbf {\bibinfo {volume}
  {98}},\ \bibinfo {pages} {032506} (\bibinfo {year} {2018})}\BibitemShut
  {NoStop}%
\end{thebibliography}%

\end{document}